\documentclass[12pt]{JHEP3}
\usepackage{latexsym}
\usepackage{epsfig}
\usepackage{amssymb}
\usepackage{amsthm}
\usepackage{amsmath}
\usepackage{amssymb,amsfonts}
\usepackage{eufrak}
\usepackage{verbatim}

\DeclareMathOperator{\diag}{\textrm{diag}}

\DeclareMathOperator{\Tr}{\textrm{Tr}}
\DeclareMathOperator{\Exp}{\textrm{Exp}}
\DeclareMathOperator{\ii}{\textrm{i}}
\DeclareMathOperator{\Imag}{\textrm{Im}}

\DeclareMathOperator{\ML}{\mathcal{L}}
\DeclareMathOperator{\MM}{\mathcal{M}}
\DeclareMathOperator{\MN}{\mathcal{N}}

\newcommand{\be}{\begin{equation}}
\newcommand{\ee}{\end{equation}}
\newcommand{\bea}{\begin{eqnarray}}
\newcommand{\eea}{\end{eqnarray}}
\renewcommand{\d}{\partial}
\newcommand{\nn}{\nonumber}

\title{Scalar-Induced Compactifications in Higher Dimensional Supergravities }

\author{
Alex Kehagias and Constantina Mattheopoulou\\
    Department of Physics, National Technical University of Athens, \\
    GR-15773 Zografou, Athens,
    Greece\\
    }

\preprint{\hepth{0507010}}

\abstract{ We discuss compactifications of higher dimensional supergravities which are induced by scalars.
In particular, we consider vector multiplets coupled to the supergravity multiplet in the case of $D=9$,$8$ and
$D=7$ minimal supergravities. These vector multiplets contain scalars, which parametrize coset spaces of the general
form~$SO(10\!-\!D,n)/SO(10\!-\!D)\!\times\!SO(n)$, where $n$ is the number of vector multiplets. We discuss the
compactification of the supergravity theory to $D\!-\!2$ dimensons, which is induced by non-trivial vacuum scalar
field configurations. There are singular and non-singular solutions, which preserve half of the supersymmetries. }

\begin{document}

\section{Introduction}
\label{sec-1}

A central issue in all higher-dimensional theories is compactification. Various mechanisms have been proposed and the general idea is to construct vacuum spacetimes of $M^4\times X$ geometry, where $M^4$ is 4D Minkowski spacetime and $X$ is a compact internal space. In string theory for example, one way of constructing appropriate vacua is to look for classical supergravity solutions. As supergravity is the low-energy limit of
string theory, supergravity solutions describe accordingly low-energy string vacua. There will be $\alpha'$--corrections as well as string-loop corrections to these solutions but, nevertheless, these solutions will still be valid in some appropriate limits. These vacua are constructed by solving the classical field equations with appropriate fields turned on. Usually such fields are antisymmetric $p$--forms as well as various scalars like the dilaton, axion etc.,  which appear in almost all supergravity theories.
%Some of these classical solutions are
%interpreted  as the D-branes of string theory. There are also
%string vacua where the are no other fields, except the graviton,
%turned on.  Such vacua are necessarily Ricci-flat
%\be
%R_{MN}=0\, , \label{Einstein} \ee i.e., they satisfy  vacuum
%Einstein equations. Solutions to the above equations with
%four-dimensional Poincar\'e invariance are provided by
%$M^4\!\times\! X$ where $M^4$ is ordinary Minkowski space-time and
%$X$ is a Ricci-flat manifold.  Supersymmetry demands that $X$
%should be a manifold  of $U(1)^6, ~SU(2)\times U(1)^2$ or $SU(3)$
%holonomy. Manifolds with such holonomies can be either compact or
%non-compact. We recall for example the case of the compact $K3$
%surface and the non-compact
% Eguchi-Hanson gravitational instanton,   both of $SU(2)$ holonomy.
The four-dimensional Plank mass $M_{P}$ is proportional to the volume $V(X)$ of $X$ \be M_{P}^2= M_s^8\,  V(X)\, , \label{Mp} \ee where $M_s^2\sim 1/\alpha'$ is the string-mass scale. Propagating gravity therefore exists in four dimensions if the volume of $X$ is finite. This is always the case for a smooth compact space $X$.

It should be stressed, however, that
%there are
%also non-compact spaces of finite volume which, according to
%eq.(\ref{Mp}) will lead to a four-dimensional dynamical gravity.
non-compact spaces may also be employed. Such spaces have been considered in the Kaluza-Klein programme
\cite{Gell-Mann1},\cite{Gell-Mann2},\cite{Nicolai}.
%The drawback of non-compact
%spaces of finite volume is that they suffer from singularities.
%However, although singularities are considered in general to be
%disastrous, there exist singularities which are quite mild in the
%sense that they can be attributed to some  form of matter. These
%are delta-function singularities which may be interpreted as
%strings, domain walls or fundamental branes in general.
%Supergravity solutions with a bulk cosmological constant and such
%singularities have been constructed in \cite{KK} Our aim here is
%to solve eq.(\ref{Einstein}) with a non-compact internal space $X$
%of finite volume and delta-functions singularities.
Adopting the proposal of a non-compact internal space, we are facing a new problem. A smooth non-compact space has infinite volume so that the four-dimensional Planck mass $M_{P}^2$ will be infinite. As a result, gravitational interactions will be actually higher-dimensional and not four-dimensional as we want. The
solution here is to assume that the non-compact space has finite volume. In this case we expect singularities and several pathologies like continuous spectra,  violation of conservation laws for energy, momentum, angular momentum etc. caused by possible leakage from the singular points. Thus, in order for our proposal to be viable, all these pathologies should be avoided.

This is the case, for example, in the tear-drop solution~\cite{Gell-Mann1},\cite{Gell-Mann2}. In this solution, the scalars of the type IIB supergravity are non-vanishing and the 10D spacetime is compactified to a space
diffeomorphic to the $\frac{SU(1,1)}{U(1)}$ scalar manifold. This compactification is triggered by a non-trivial scalar field configuration. The spacetime develops a naked singularity which, however, is harmless and does not lead to any physically unacceptable situation as all physical quantities, like energy, momentum and angular momentum are conserved. It should be noted that a similar solution is also the stringy cosmic string where, in addition, the non-perturbative $SL(2,\mathbb{Z})$ symmetry of type IIB is employed for the compactification \cite{GV}.

Here, we will show that tear-drop-like solutions are actually quite generic, using $D=9,8,7$ minimal supergravities as a concrete example. The supergravities under consideration contain numerous scalar fields, combinations of which may trigger compactifications of the sort described in the preceding paragraph. Adopting a convenient parameterization of the scalar sector of these theories in terms of solvable Lie algebras, we can write the scalar Lagrangian and the equations of motion of the theory in a compact form. We can then identify solutions of the equations of motion, in which the non-vanishing scalars form $\frac{SL(2,\mathbb{R})}{U(1)}$ submanifolds of the original scalar manifold, inducing tear-drop-like compactifications. To be specific, there are singular and non-singular solutions. The former are like the tear-drop~\cite{Gell-Mann1},\cite{Gell-Mann2}, while the latter are like the stringy cosmic string~\cite{GV}. Moreover, in the case of $D=8$ supergravity, there is also a compactification to 4D, achieved by two vector multiplets coupled to the supergravity multiplet. The compactifications found preserve half the supersymmetries of the original theory in all cases.

In the following section 2, in order to set up our notation, we introduce an appropriate parametrization for the $\MM$ spaces. In section 3, we briefly review the $D=9,8,7$ minimal supergravities and, after deriving the field equations with scalar fields turned on, we present their solutions. In section 4, we discuss the
supersymmetric properties of the solutions. Finally, in section 5, we comment on our findings.

\section{Scalar Coset Manifolds in Minimal Supergravities}
\label{sec-2}

The scalar fields of minimal supergravities coupled to matter in $D=9,8,7,5,4$ (apart from the dilaton contained in the supergravity multiplet), parameterize non-compact coset manifolds of $\frac{ SO(10-D,n)}{SO(10-D) \times SO(n)}$ type\footnote{In the special case $D=5$, there are actually more possibilities for the scalar manifold.}. To see how these manifolds arise, we first note that in these theories the supergravity multiplet contains $10-D$ vector fields while the vector multiplet contains $10-D$ scalars. These two types of fields carry indices of the R-symmetry group of the supersymmetry algebra which, for the specific dimensions, is isomorphic to $SO(10-D)$. By combining the supergravity multiplet with $n$ vector multiplets, the total $10-D+n$ vectors fall into the defining representation of $SO(10-D,n)$, so that the latter is identified as a global symmetry group of the theory. On the other hand, the theory describing the $n(10-D)$ real scalars from the vector multiplets is invariant under $SO(n)$ rotations between the scalars of the same R-symmetry index in different multiplets as well as under $SO(10-D)$ R-symmetry rotations within each vector multiplet. Therefore, the scalar manifold, i.e. the space of inequivalent points parameterized by the scalars, is the coset space $\MM=\frac{SO(10-D,n)}{SO(10-D) \times SO(n)}$. The scalar isometry group $SO(10-D,n)$ is a non-compact real form of $D_\ell$ (for $10-D+n=2\ell$) or of $B_\ell$ (for $10-D+n=2\ell+1$).

Here, we will first review the parameterizations of scalar cosets in both the coset-manifold and the group-manifold approaches. Using the second approach, we will then give the general form of the scalar Lagrangian. Finally, we will specialize to the $\frac{SO(10-D,n)}{SO(10-D) \times SO(n)}$ coset spaces for the
dimensions $D=9,8,7$ and we will discuss the further simplifications that occur in the structure of the scalar
Lagrangian.

\subsection{The coset and the group manifold approaches}
\label{sec-2-1}

Consider a general non-compact coset manifold $\MM=G/H$, where $G$ is some group associated with a non-compact real form $\mathfrak{g}_{nc}$ of a complex Lie algebra $\mathfrak{g}$ and $H$ is its subgroup associated with the maximal compact subalgebra $\mathfrak{h}$ of $\mathfrak{g}_{nc}$. Such a manifold admits two different descriptions, which we will outline below.

The first is the usual coset-manifold description, based on the Cartan decomposition $\mathfrak{g}_{nc} = \mathfrak{h} \oplus \mathfrak{k}$, where $\mathfrak{k}$ is the non-compact complementary subspace of $\mathfrak{h}$ within $\mathfrak{g}_{nc}$; this subspace is not an algebra. In this description, a coset representative is defined as $L_K=\Exp(k)$ with $k \in \mathfrak{k}$. Since $\mathfrak{k}$ is not an algebra,
$L_K$ is not a group element but, in general, includes an $H$--valued part. This description is the usual one employed in the construction of supergravity theories. In particular, the decomposition of the Maurer-Cartan form of a coset representative into an $\mathfrak{h}$--valued and a $\mathfrak{k}$--valued part gives the composite connections and the composite coset vielbeins respectively in a form that may be directly used in the supersymmetry transformation laws. However, the explicit form of the scalar Lagrangian in this description is quite complicated.

The second description is the group-manifold description, which is based on solvable Lie algebras\cite{Gilmore,Helgason}. This description is based on the Iwasawa decomposition\cite{Helgason} which ensures that, for any non-compact real form $\mathfrak{g}_{nc}$ of $\mathfrak{g}$, there exists a solvable Lie algebra $\mathrm{Solv}(\mathfrak{g}_{nc})$ such that $\mathfrak{g}_{nc}$ may be decomposed as the direct sum $\mathfrak{g}_{nc} = \mathfrak{h} \oplus \mathrm{Solv}(\mathfrak{g}_{nc})$. The solvable Lie algebra is constructed as follows. We first consider a Cartan-Weyl basis for the generators of $\mathfrak{g}$, denoting the $\ell$ generators of the Cartan subalgebra $\mathfrak{a}$ by $\{H_I\}$ and the positive-root generators by $\{E_A\}$; the set of the positive roots is denoted as $\Phi^+$. The solvable Lie algebra is then given by the direct sum
\be
\label{2-1}
\mathrm{Solv}(\mathfrak{g}_{nc}) = \mathfrak{a}_{nc} \oplus \mathfrak{n}.
\ee
Here $\mathfrak{a}_{nc}$ is the non-compact part of the Cartan subalgebra $\mathfrak{a}$ of $\mathfrak{g}$,
\be
\label{2-2}
\mathfrak{a}_{nc} = \mathfrak{a} \cap \mathfrak{k},
\ee
generated by an appropriate subset $\{ H_i \}$ of the Cartan generators and $\mathfrak{n}$ is the algebra which is constructed from the set $\{ E_\alpha \}$ of the positive-root generators of $\mathfrak{g}$ that do not commute with all $H_i$'s according to the relation
\be
\label{2-3}
\mathfrak{n} = (
\sum_{\alpha \in \Delta^+} E_\alpha ) \cap \mathfrak{g}_{nc},
\ee
where $\Delta^+$ is the subset of $\Phi^+$ containing the positive roots associated with $\{ E_\alpha \}$.
The intersection symbol in (\ref{2-3}) denotes that the $\{ E_\alpha \}$ should be arranged in suitable linear combinations that belong to the non-compact real form $\mathfrak{g}_{nc}$. We note that, in the special case where $G$ is a split group (i.e. where $\MM=G/H$ is maximally non-compact), the solvable Lie algebra of
$\mathfrak{g}_{nc}$ coincides with its Borel subalgebra generated by all $H_I$'s and all $E_A$'s; in the case of $G=SO(10-D,n)$, this happens only when $n = 10 - D$ or $n = 10 - D \pm 1$. In the group-manifold description, a coset representative is defined as $L_S=\Exp(s)$ with $s \in \mathrm{Solv}(\mathfrak{g}_{nc})$ and, unlike the case in the coset-manifold description, it \emph{is} an element of $\mathrm{Solv}(\mathfrak{g}_{nc})$, since the latter is an algebra (for a given point $P \in \MM$, the representatives $L_S(P)$ and $L_K(P)$ are equivalent up to a right-multiplication by an element of $H$). Specifically, the Iwasawa decomposition implies that a general element $\mathbf{g} \in G$ can be uniquely
decomposed as
\be
\label{2-4} \mathbf{g} = \Exp(h) \Exp(a) \Exp(n); \qquad h \in \mathfrak{h},a \in \mathfrak{a}_{nc},n \in
\mathfrak{n}.
\ee
As a result, $\mathbf{g}$ can be written  as the product of elements obtained by exponentiation of the maximal compact subalgebra $\mathfrak{h}$, the non-compact Cartan subalgebra $\mathfrak{a}_{nc}$ and the subalgebra $\mathfrak{n}$ associated with the positive roots in $\Delta^+$. According to this decomposition, a coset
representative is obtained from (\ref{2-4}) by discarding the $H$--valued factor and is thus given by
\be
\label{2-5}
L_S = \Exp(a) \Exp(n); \qquad a \in \mathfrak{a}_{nc},n \in
\mathfrak{n}. \ee
The advantage of the group-manifold description of the scalar coset in supergravity theories is that, unlike the
coset-manifold description, it leads to a natural one-to-one correspondence between the scalar fields of the theory with the generators $\{H_i\}$ and $\{E_\alpha\}$ which form an algebra. For that reason, the explicit form of the scalar coset Lagrangian simplifies considerably through the use of group-theoretical methods. The application of solvable Lie algebras in supergravity has been extensively studied in \cite{Andrianopoli:1996bq},\cite{Andrianopoli:1996zg},\cite{Arcioni:1998mn},\cite{Castellani:1998nz},\cite{Lu:1998xt}.

\subsection{The scalar Lagrangian}
\label{sec-2-2}

Here, we will review the construction of scalar coset sigma-model Lagrangians using the group-manifold approach; the detailed procedure can be found e.g. in \cite{Keurentjes:2002xc},\cite{Keurentjes:2002rc},\cite{Yilmaz:2003as},\cite{Dereli:2004uh}.  According to the remarks of the previous paragraph, one may use
the Iwasawa decomposition (\ref{2-5}) to parameterize a coset representative. In the context of a sigma model, the Lie-algebra-valued quantities $a \in \mathfrak{a}_{nc}$ and $n \in \mathfrak{n}$ are taken to be functions of the spacetime coordinates $x^M$. They can be expressed in terms of the fields $\{ \phi^i(x) \}$ (dilatons,  corresponding to $\{ H_i \}$) and $\{ \chi^\alpha(x) \}$ (axions, corresponding to $\{ E_\alpha \}$) as
follows
\be
\label{2-6}
a(x) = \frac{1}{2} \phi^i(x) H_i ,\qquad n(x) = \chi^\alpha(x) E_\alpha.
\ee
So, the Iwasawa decomposition reads
\be
\label{2-7}
L = e^{\frac{1}{2} \phi^i H_i} e^{\chi^\alpha E_\alpha}.
\ee
We stress again that the root-space generators $\{ E_\alpha \}$ are in fact restricted to enter (\ref{2-6}) and (\ref{2-7}) only through the appropriate linear combinations that belong to the chosen real form of the isometry algebra. This does not alter at all the discussion that follows.

The Lagrangian of the sigma model coupled to gravity is expressed in terms of $L$ by
\be
\label{2-8}
e^{-1} \ML_s = - \frac{1}{4} \Tr \left[ (\partial_M L L^{-1}) ( \partial^M L L^{-1} )^\# + (
\partial_M L L^{-1}) (\partial^M L L^{-1}) \right],
\ee
where, ``$\#$'' denotes the generalized transpose, defined as the transformation induced by the Cartan involution on the group elements. It is easily shown \cite{Keurentjes:2002xc} that the Maurer-Cartan form appearing in (\ref{2-8}) has the explicit form
\be \label{2-9}
\partial_M L L^{-1} = \frac{1}{2} \partial_M \phi^i H_i + e^{\frac{1}{2} \alpha_i \phi^i} F^\alpha_M
E_\alpha, \ee
with $\alpha_i$  the $i$-the component of the root $\alpha$. Also, $F^\alpha_M$ is the
field strength associated with $\chi^\alpha$, given by \cite{Yilmaz:2003as}
\be
\label{2-10}
F^\alpha_M = \partial_M \chi^\alpha + \frac{1}{2!} ( \chi^\gamma C^\alpha_{\gamma\beta} ) \partial_M \chi^\beta +
\frac{1}{3!} ( \chi^\gamma C^\alpha_{\gamma\varepsilon} ) ( \chi^\delta C^\varepsilon_{\delta\beta} ) \partial_M
\chi^\beta + \ldots
\ee
where $C^\gamma_{\alpha\beta}$ are the structure constants in  $[E_\alpha,E_\beta] = C^\gamma_{\alpha\beta} E_\gamma$, i.e $C^\gamma_{\alpha\beta} = N_{\alpha,\beta}$ if $\alpha+\beta=\gamma$ and zero otherwise. Using the parameterization (\ref{2-9}), one can show that the scalar Lagrangian takes the simple form
\be
\label{2-11} e^{-1} \ML_s = - \frac{1}{4} \sum_i (\partial_M
\phi^i)^2 - \frac{1}{2} \sum_{\alpha} e^{\alpha_i \phi^i}
(F^\alpha_M)^2,
\ee
where we employed a normalization with $\Tr H_i H_j = 2 \delta_{ij}$ and $\Tr E_\alpha E_{-\alpha}=2$.

\subsection{ The $\frac{SO(10-D,n)}{SO(10-D) \times SO(n)}$ scalar coset}
\label{sec-2-3}

After this general discussion, let us specialize to the $\frac{SO(10-D,n)}{SO(10-D) \times SO(n)}$
case of interest; here we will always assume that $n \geqslant 10-D$. To construct the $D_\ell$ or $B_\ell$ root vectors associated with $SO(10-D,n)$, we define $\epsilon_i$ as the $\ell$--dimensional vector whose $i$--th
element is unity with all other elements zero. Then, the positive roots of $D_\ell$ or $B_\ell$ are given by \bea
\label{2-12}
\epsilon_i \pm \epsilon_j &;& \qquad i<j=1,\ldots \ell, \nn \\
\epsilon_i &;& \qquad i=1,\ldots \ell \quad \text{(only for $B_\ell$)}.
\eea
and so the associated generators are $E_{\epsilon_i \pm \epsilon_j}$ and $E_{\epsilon_i}$.

To construct the solvable Lie algebra, we choose our conventions so that the non-compact Cartan generators $\{ H_i \}$ of $\mathfrak{a}_{nc}$ are given by $\{ H_{\epsilon_i} \}$, $i=1,\ldots,10-D$. The generators of the nilpotent subalgebra $\mathfrak{n}$ are then found by considering the subset of the generators $\{ E_{\epsilon_i \pm \epsilon_j}, E_{\epsilon_i} \}$ that do not commute with all of the $H_i$'s and taking
appropriate linear combinations that belong to the non-compact real form of $D_\ell$ or $B_\ell$ appropriate for the space of interest. Explicit constructions will be shown below.

In what follows, we will further specialize to the cases of interest, namely $D=9,8,7$. In particular, we will apply the above procedure to construct the solvable Lie algebra and then, examining its structure, we will see how the scalar Lagrangian simplifies.
\begin{itemize}

\item $D=9$. In the $D=9$ case, corresponding to $\frac{SO(1,n)}{ SO(n)}$, there is only one non-compact
Cartan generator which, in our conventions is taken to be
\be \label{2-13} H_{\epsilon_1}.
\ee
The generators of the solvable algebra are then given by the positive-root generators~\cite{Lu:1998xt},
which already belong to the $\mathfrak{so}(1,n)$ real form
\bea
\label{2-14}
&  E_{\epsilon_1 + \epsilon_i}  ,\qquad E_{\epsilon_1 - \epsilon_i} ; \qquad i=2,\ldots \ell \nn\\
&  E_{\epsilon_1},  \qquad \text{(only for $B_\ell$)},
\eea

In this case, the structure of the scalar Lagrangian is very simple. Indeed, all structure constants associated with positive-root generators vanish,
\be
\label{2-15} C^\alpha_{\beta\gamma} = 0,
\ee
and so the field strengths (\ref{2-10}) are simply $F^\alpha_M = \partial_M \chi^\alpha$.

\item $D=8$. In the $D=8$ case, the scalar coset manifold is $\frac{SO(2,n)}{ SO(2) \times SO(n)}$, with $n \geqslant 2$. There are two non-compact Cartan generators which can be chosen as
\be
\label{2-16} H_{\epsilon_1}, \qquad H_{\epsilon_2}.
\ee
The generators of the solvable algebra are then given by the positive-root generators
\bea
\label{2-17}
&E_{\epsilon_1 + \epsilon_2} ,\qquad E_{\epsilon_1 - \epsilon_2}, \nn\\ & E_{\epsilon_1},\qquad  E_{\epsilon_2}, \qquad \text{(only for
$B_\ell$)},
\eea
that are already in the $\mathfrak{so}(2,n)$ real form, plus the linear combinations \cite{Castellani:1998nz},\cite{Lu:1998xt}
\bea
\label{2-18}
& \frac{1}{\sqrt{2}}( E_{\epsilon_1 + \epsilon_i} + E_{\epsilon_1 - \epsilon_i} ) ,\qquad - \frac{\ii}{\sqrt{2}}(E_{\epsilon_1 + \epsilon_i} - E_{\epsilon_1 - \epsilon_i})  ; \qquad i=3,\ldots \ell, \nn\\
& \frac{1}{\sqrt{2}}( E_{\epsilon_2 + \epsilon_i} + E_{\epsilon_2 - \epsilon_i} ) ,\qquad
- \frac{\ii}{\sqrt{2}}(E_{\epsilon_2 + \epsilon_i} - E_{\epsilon_2 - \epsilon_i})
; \qquad i=3,\ldots \ell. \eea

To see the simplifications that occur regarding the coset Lagrangian, we are finding
that the only nonzero structure constants $C^\alpha_{\beta\gamma}$(\ref{2-10}) are given by
\bea \label{2-19} & C^{\epsilon_1 \pm \epsilon_i}_{\epsilon_1 - \epsilon_2,\epsilon_2 \pm
\epsilon_i} , \quad C^{\epsilon_1 + \epsilon_2}_{\epsilon_1 \pm \epsilon_i,\epsilon_2 \mp \epsilon_i} \nn\\ & C^{\epsilon_1 + \epsilon_2}_{\epsilon_1, \epsilon_2} , \quad C^{\epsilon_1}_{\epsilon_1- \epsilon_2, \epsilon_2} \qquad \text{(only for $B_\ell$)}.
\eea
It follows then that the only nonzero terms of Eq. (\ref{2-10}) quadratic in the structure constants involve the combinations
\bea
\label{2-20}
&C^{\epsilon_1 + \epsilon_2}_{\epsilon_2 \pm \epsilon_i,\epsilon_1 \mp \epsilon_i} C^{\epsilon_1 \mp \epsilon_i}_{\epsilon_1 - \epsilon_2,\epsilon_2 \mp \epsilon_i}, \quad C^{\epsilon_1 + \epsilon_2}_{\epsilon_2
\mp \epsilon_i,\epsilon_1 \pm \epsilon_i} C^{\epsilon_1 \pm \epsilon_i}_{\epsilon_2 \pm \epsilon_i,\epsilon_1 - \epsilon_2}, \nn\\
&C^{\epsilon_1 + \epsilon_2}_{\epsilon_2,\epsilon_1}
C^{\epsilon_1}_{\epsilon_1-\epsilon_2,\epsilon_2}, \quad
C^{\epsilon_1 + \epsilon_2}_{\epsilon_2,\epsilon_1}
C^{\epsilon_1}_{\epsilon_2,\epsilon_1-\epsilon_2} \qquad
\text{(only for $B_\ell$)},
\eea
and that terms of cubic or higher order vanish since there are no available indices to contract; this implies that, in this case, Eq. (\ref{2-10}) actually contains no further terms. The important fact that we will need later on is that there are no nonzero structure constants with $\epsilon_1+\epsilon_2$ as a lower index.

\item $D=7$. In the $D=7$ case, the scalar coset manifold is $\frac{SO(3,n)}{SO(3) \times SO(n)}$, with
$n \geqslant 3$. There are three non-compact Cartan generators which can be chosen as
\be
\label{2-21}
H_{\epsilon_1}, \qquad H_{\epsilon_2} ,\qquad H_{\epsilon_3}.
\ee
In a similar manner as before, the solvable generators are given by the positive-root generators
\bea
\label{2-22}
&E_{\epsilon_1 + \epsilon_2},\quad E_{\epsilon_1 - \epsilon_2},
\quad E_{\epsilon_1 + \epsilon_3},\quad E_{\epsilon_1 - \epsilon_3},
\quad E_{\epsilon_2 + \epsilon_3},\quad E_{\epsilon_2 - \epsilon_3}, \nn\\
& E_{\epsilon_1},\quad   E_{\epsilon_2}, \quad  E_{\epsilon_3}, \qquad \text{(only
for $B_\ell$)},
\eea
and the combinations~\cite{Lu:1998xt}
\bea
\label{2-23}
&\frac{1}{\sqrt{2}} ( E_{\epsilon_1 + \epsilon_i} + E_{\epsilon_1 - \epsilon_i} ),\qquad - \frac{\ii}{\sqrt{2}}(E_{\epsilon_1 + \epsilon_i} -  E_{\epsilon_1 - \epsilon_i});& \qquad i=4,\ldots \ell, \nn\\
&\frac{1}{\sqrt{2}} ( E_{\epsilon_2 + \epsilon_i} + E_{\epsilon_2 - \epsilon_i} ) ,\qquad - \frac{\ii}{\sqrt{2}}(E_{\epsilon_2 + \epsilon_i} - E_{\epsilon_2 - \epsilon_i});& \qquad i=4,\ldots \ell, \nn\\
&\frac{1}{\sqrt{2}} ( E_{\epsilon_3 + \epsilon_i} + E_{\epsilon_3 - \epsilon_i} ) ,\qquad - \frac{\ii}{\sqrt{2}}(E_{\epsilon_3 + \epsilon_i} - E_{\epsilon_3 - \epsilon_i});& \qquad i=4,\ldots \ell.
\eea

In this case, the structure of the Lagrangian is more complicated since there exist structure-constant combinations of cubic and quartic (in the $D_\ell$ case) order. As in the previous case, no nonzero structure constants with an $\epsilon_1+\epsilon_2$ lower index arise.
\end{itemize}

\section{Scalar-induced compactifications in $D=9,8,7$ minimal supergravities}
\label{sec-3-1}

As stated in the intruduction, the mechanism of scalar-induced compactification admits generalizations to supergravities of diverse dimensions. In this section, we demonstrate the existence of such solutions for the case of the minimal supergravities in $D=9,8,7$. We start by discussing the basic aspects of these minimal supergravities. Next, we use the parameterization of the previous section to write down the field equations for a vacuum configuration containing only gravity and scalars. Finally, we show that these equations are consistent with a particular ansatz for the scalars and we present the resulting solutions.
Having described the basic aspects of the scalar coset manifolds, w will describe next the general aspects of the $D=9,8,7$ supergravities of interest.

\subsection{Minimal supergravities in $D=9,8,7$}

\subsubsection{$D=9$ supergravity}

The field content of the massless representations of the $D=9$, $\mathcal{N}=2$ supersymmetry algebra consists of the following multiplets
\begin{eqnarray}
\text{Supergravity multiplet} \quad&:&\quad ( g_{MN}, B_{MN}, A_M , \sigma, \psi_M, \chi ), \nonumber\\
\text{Vector multiplet} \quad&:&\quad ( A_M, \varphi, \lambda ).
\label{e-3-1}
\end{eqnarray}
where all spinors are pseudoMajorana. A general $D=9$, $\MN=2$ supergravity theory is constructed by combining the supergravity multiplet with $n$ vector multiplets. This leads to the reducible multiplet
\be
(g_{MN}, B_{MN}, A^I_M, \varphi^{\bar{\alpha}},
\sigma, \psi_M, \chi, \lambda^{\bar{a}} ).
\label{e-3-2}
\ee
where $\bar{\alpha}=1,\ldots,n$ labels the scalars, $\bar{a}=1,\ldots,n$ labels the gauginos and $I=1,\ldots,n+1$ labels the vectors (here, we employ barred indices, e.g. $\bar{\alpha}$, in order to avoid
confusion with the indices appearing in section 2).

As mentioned before, the $n$ scalars $\varphi^{\bar{\alpha}}$ parameterize the non-compact coset manifold $\frac{SO(1,n)}{SO(n)}$ (more precisely, they parameterize the space $H^n = \frac{SO_0 (1,n)}{SO(n)}$, which is the upper sheet of a hyperboloid). To parameterize the scalar coset, one introduces a coset representative $L=\{L_I^{\phantom{I} A}\}$ given by $(n+1) \times (n+1)$ matrix in the vector representation\footnote{In the coset-manifold approach, the usual choice is to combine the $n$ scalars into a column vector $\Phi$ and define $L = \exp \left( \begin{array}{cc} 0 & \Phi \\ \Phi^T & 0 \end{array} \right)$.} (here, $A$ is the curved index analog to $I$). The inverse matrix, given by $L^{-1} = \{L_A^{\phantom{A} I}\}$,
satisfies
\be
L_A^{\phantom{A} I} L_I^{\phantom{I} B} = \delta_A^{\phantom{A} B}. \label{e-3-3}
\ee
The elements of $L$ and its inverse can be decomposed as
\be
L_I^{\phantom{I} A} = \left( L_I , L_{I}^{\phantom{I} \bar{a}} \right) ,\qquad
L_A^{\phantom{A} I} = \left( L^I , L_{\bar{a}}^{\phantom{\bar{a}}
I} \right). \label{e-3-4}
\ee
Also, $L$ satisfies the $SO(1,n)$ orthogonality condition
\be
\label{3-5}
\eta_{AB} L_I^{\phantom{I} A} L_J^{\phantom{J} B} = \eta_{IJ},
\ee
where $\eta_{AB} = \eta_{IJ} = \diag (-1,+1,\ldots,+1)$ is the $SO(1,n)$ invariant tensor. A quantity of interest constructed out of $L$ is the tensor
\be
\label{e-3-6} a_{IJ} = L_I L_J + L_I^{\phantom{I} \bar{a}} L_{J}^{\phantom{J} \bar{a}},
\ee
which contracts $I$ and $J$ so as to yield a ghost-free kinetic term for the vectors.

To construct scalar kinetic terms and covariant derivatives in the coset-manifold approach, one considers the Maurer-Cartan form\footnote{Note that, due to the particular parameterization of the coset, we have to use right-invariant Maurer-Cartan forms instead of the left-invariant ones usually employed in the supergravity literature.} of $L$, given by $\partial_M L L^{-1}$. This matrix-valued one-form decomposes into the coset vielbein $P_{M}^{\phantom{M} \bar{a}}$ and the $SO(n)$ composite connection $Q_{M \bar{a}}^{\phantom{M \bar{a}} \bar{b}}$. In the standard vector representation of $SO(1,n)$, this decomposition has the form
\be
\label{3-7}
\partial_M L L^{-1} = \left( \begin{array}{cc} 0 & P_M^{\phantom{M} \bar{a}} \\ P_{M \bar{a}}
& Q_{M \bar{a}}^{\phantom{M \bar{a}} \bar{b}} \end{array} \right) ,
\ee

The Lagrangian of the theory was first constructed in \cite{Gates:1984kr}. Its bosonic part is given by
\bea
e^{-1} \mathcal{L} &=& \frac{1}{2} R - \frac{1}{12} e^{2\sigma} G_{MNP}^2 - \frac{1}{4}
e^{\sigma} a_{IJ} F^I_{MN} F^{J MN} - \frac{7}{16} ( \partial_M \sigma )^2 - \frac{1}{4} P_M^{\phantom{M} \bar{a}} P^M_{\phantom{M} \bar{a}} \label{e-3-8}
\eea
where the field strengths $G_{MNP}$ and $F^I_{MN}$ are defined as
\be
\label{3-9}
G_{MNP} = 3 \left( \partial_{[M} B_{NP]} + \eta_{IJ} F^I_{[ MN} A^J_{P]} \right) ,\qquad F^I_{MN} = 2
\partial_{[M} A^I_{N]}. \ee

\subsubsection{$D=8$ supergravity}

The massless representations of the $D=8$, $\mathcal{N}=1$ supersymmetry algebra form the following multiplets \bea
\label{3-10}
\text{Supergravity multiplet} \quad&:&\quad ( g_{MN}, B_{MN}, A^{\bar{i}}_M , \sigma, \psi_M, \chi ), \nonumber\\
\text{Vector multiplet} \quad&:&\quad ( A_M, \varphi^{\bar{i}},
\lambda ).
\eea
where all spinors are pseudoMajorana and the index $\bar{i}=1,2$ refers to the $SO(2)$ R-symmetry group. A general $D=8$, $\mathcal{N}=1$ supergravity theory is constructed by combining the supergravity multiplet with $n$ vector multiplets. This leads to the reducible multiplet
\be \label{3-11} (g_{MN}, B_{MN}, A_M^I, \varphi^{\bar{\alpha}}, \sigma, \psi_M, \chi,
\lambda^{\bar{a}} ). \ee
where $\bar{\alpha}=1,\ldots,2n$ labels the scalars, $\bar{a}=1,\ldots,n$ labels the gauginos and
$I=1,\ldots,n+2$ labels the vectors.

The $2n$ scalars $\varphi^{\bar{\alpha}}$ parameterize the non-compact coset manifold $\frac{SO(2,n)}{ SO(2) \times SO(n)}$. The coset representative $L$ and its inverse, now given by $(n+2) \times (n+2)$ matrices, are defined in a similar manner as before, they can be decomposed as
\be
L_I^{\phantom{I} A} = \left( L_{I}^{\phantom{I} \bar{i}}, L_{I}^{\phantom{I} \bar{a}} \right)
,\qquad L_A^{\phantom{A} I} = \left( L_{\bar{i}}^{\phantom{\bar{i}} I} , L_{\bar{a}}^{\phantom{\bar{a}}
I} \right). \label{e-3-12}
\ee
They satisfy the orthogonality condition
\be
\label{3-13} \eta_{AB} L_I^{\phantom{I} A} L_J^{\phantom{J} B} = \eta_{IJ},
\ee
where $\eta_{AB} = \eta_{IJ} = \diag (-1,-1,+1,\ldots,+1)$ is the $SO(2,n)$ invariant tensor. The tensor needed for the contraction of $I$ and $J$ in the vector kinetic term is now
\be
\label{3-14} a_{IJ} = L_I^{\phantom{I} \bar{i}} L_J^{\phantom{J} \bar{i}} +
L_I^{\phantom{I} \bar{a}} L_J^{\phantom{J} \bar{a}}
\ee

To proceed, we consider the Maurer-Cartan form of $L$, which contains the coset vielbein $P_{M \bar{i}}^{\phantom{M \bar{i}} \bar{a}}$ and the $SO(2)$ and $SO(n)$ composite connections $Q_{M
\bar{i}}^{\phantom{M \bar{i}} \bar{j}}$ and $Q_{M \bar{a}}^{\phantom{M \bar{a}} \bar{b}}$. In the vector
representation, the decomposition has the form
\be
\label{3-15}
\partial_M L L^{-1} = \left( \begin{array}{cc} Q_{M
{\bar{i}}}^{\phantom{M {\bar{i}}} {\bar{j}}} & P_{M
\phantom{\bar{a}} \bar{i}}^{\phantom{M} \bar{a}} \\ P_{M
\bar{a}}^{\phantom{M \bar{a}} {\bar{i}}} & Q_{M
\bar{a}}^{\phantom{M \bar{a}} \bar{b}} \end{array} \right) ,
\ee
For later convenience, it is also useful to define the quantities
\be \label{3-16}
\hat{P}_{M \bar{a}} \equiv  P_{M \bar{a}}^{\phantom{M \bar{a}} 1} + i \gamma_{(9)} P_{M
\bar{a}}^{\phantom{M \bar{a}} 2} , \qquad Q_M \equiv Q_{M 1}^{\phantom{M 1} 2}.
\ee

The Lagrangian of the theory was first constructed in \cite{Salam:1985ns}. Its bosonic part is given by
\be
\label{3-17}
e^{-1} \ML = \frac{1}{2} R - \frac{1}{12} e^{2 \sigma} G_{MNP}^2 - \frac{1}{4} e^\sigma
a_{IJ} F^I_{MN} F^{J MN}  - \frac{3}{8} (\partial_M \sigma )^2 - \frac{1}{2} P_{M \phantom{\bar{a}}
\bar{i}}^{\phantom{M} \bar{a}} P^{M \phantom{\bar{a}} \bar{i}}_{\phantom{M}\bar{a}},
\ee
with the field strengths $G_{MNP}$ and $F^I_{MN}$ defined as in (\ref{3-9}).

\subsubsection{$D=7$ supergravity}

The field content of the massless representations of the $D=7$, $\mathcal{N}=2$ supersymmetry algebra consists of the following multiplets
\bea
\label{3-18}
\text{Supergravity multiplet} \quad&:&\quad ( g_{MN}, B_{MN}, A^{\phantom{M}
\bar{i}}_{M \phantom{\bar{i}} \bar{j}}, \sigma, \psi^{\bar{i}}_M, \chi^{\bar{i}} ), \nonumber\\
\text{Vector multiplet} \quad&:&\quad ( A_M,
\varphi^{\bar{i}}_{\phantom{\bar{i}} \bar{j}}, \lambda^{\bar{i}}
).
\eea
where all spinors are symplectic Majorana and the index $\bar{i}=1,2$ labels the fundamental representation of the $Sp(1) \cong SU(2) \cong SO(3)$ R-symmetry group. A general $D=7$, $\mathcal{N}=2$ supergravity theory is constructed by combining the supergravity multiplet with $n$ vector multiplets. This leads to the reducible multiplet
\be
\label{3-19}
( g_{MN}, B_{MN}, A_M^I, \varphi^{\bar{\alpha}}, \sigma, \psi^{\bar{i}}_\mu, \chi^{\bar{i}}, \lambda^{\bar{a}\bar{i}} ),
\ee
where $\bar{\alpha}=1,\ldots,3n$ labels the scalars, $\bar{a}=1,\ldots,n$
labels the gauginos, and $I=1,\ldots,n+3$ labels the vectors resulting from the combination of
$A^{\phantom{\mu} \bar{i}}_{\mu \phantom{\bar{i}} \bar{j}}$ and $A^{\bar{a}}_\mu$. Our notation and conventions are as in~\cite{AKK},\cite{AKK2}.

The $3n$ scalars $\varphi^{\bar{\alpha}}$ parameterize the non-compact coset space $\frac{SO(3,n)}{SO(3)\times SO(n)}$. Its representative $L$ and its inverse are $(n+3) \times (n+3)$ matrices, they are decomposed as
\be
\label{3-20}
L_I^{\phantom{I} A} = \left( L_{I \phantom{\bar{i}} \bar{j}}^{\phantom{I} \bar{i}}, L_I^{\phantom{I} \bar{a}} \right) ,\qquad L_A^{\phantom{A} I} = \left( L_{\bar{i}}^{\phantom{\bar{i}} \bar{j} I},
L_{\bar{a}}^{\phantom{\bar{a}} I} \right),
\ee
and they satisfy
\be
\label{3-21}
\eta_{AB} L_I^{\phantom{I} A} L_J^{\phantom{J} B} = \eta_{IJ},
\ee
with $\eta_{AB} = \eta_{IJ} = \diag(-1,-1,-1,+1,\ldots,+1)$.

The Maurer-Cartan form of $L$ contains the coset vielbein $P_{M \bar{a} \bar{i}}^{\phantom{M\bar{a} \bar{i}} \bar{j}}$ and the $SO(3)$ and $SO(n)$ composite connections $Q_{M \bar{i}}^{\phantom{M\bar{i}} \bar{j}}$ and $Q_{M \bar{a}}^{\phantom{M \bar{a}} \bar{b}}$. The decomposition is as follows
\be
\label{3-22}
\partial_M L L^{-1} = \left( \begin{array}{cc} Q_{M \bar{i}}^{\phantom{M\bar{i}}
\bar{j}} & P_{M \phantom{\bar{a}\bar{i}} \bar{j}}^{\phantom{M} \bar{a}\bar{i}} \\
P_{M \bar{a} \bar{i}}^{\phantom{M\bar{a} \bar{i}} \bar{j}} &
Q_{M \bar{a}}^{\phantom{M \bar{a}} \bar{b}} \end{array} \right).
%\label{e-2-9}
\ee
We note that here we have followed the usual conventions by replacing the $SO(3)$ vector index $\bar{X}=1,2,3$ naturally appearing in the vector representation,  by a pair of symmetric $SO(3)$ spinor (or $Sp(1)$ fundamental) indices $\bar{i}\bar{j}$. The transformation between the two types of notation will be shown and utilized in \S\ref{sec-4-3}.

The bosonic Lagrangian of the theory \cite{Bergshoeff:1985mr} is given by
\be
e^{-1} \mathcal{L} = \frac{1}{2} R - \frac{1}{12} e^{2 \sigma} G_{MNP}^2 -
\frac{1}{4} e^{\sigma} a_{IJ} F^I_{MN} F^{JMN} - \frac{5}{8} (\partial_M \sigma)^2 -
 \frac{1}{2} P_{M \phantom{\bar{a}\bar{i}} \bar{j}}^{\phantom{M} \bar{a}\bar{i}}
 P^{M \phantom{\bar{a}} \bar{j}}_{\phantom{M} \bar{a} \phantom{\bar{j}} \bar{i}}.
\label{e-3-23}
\ee

\subsection{Field equations}
\label{sec-3-2}

Our next task is to derive and solve the field equations, which follow from the $D=9,8,7$ dimensional Lagrangians (\ref{e-3-8},\ref{3-17},\ref{e-3-23}), respectively. We will assume a configuration where  the only nonzero fields besides gravity are the scalars $\varphi^{\bar{\alpha}}$. Employing the decomposition $\varphi^{\bar{\alpha}} = (\phi^i,\chi^\alpha)$ of section 2, and appropriately writing the scalar Lagrangian as
\be
\label{3-24}
e^{-1} \ML_s = - \frac{1}{8} \sum_{i=1}^\ell (\partial_M \phi^i)^2
- \frac{1}{4} \sum_{\alpha} e^{\alpha_i \phi^i} (F^\alpha_M)^2,
\ee
one easily sees that the non-trivial equations of motion, in $D=9,8,7$ are the Einstein equation
\be
\label{3-25} R_{MN} = \frac{1}{4} \sum_{i=1}^\ell \partial_M \phi^i \partial_N \phi^i +
\frac{1}{2} \sum_{\alpha} e^{\alpha_i \phi^i} F^\alpha_M
F^\alpha_N, \ee
the scalar equations for $\phi^i$,
\be
\label{3-26} \Box
\phi^i = \sum_{\alpha} \alpha_i e^{\alpha_i \phi^i} ( F^\alpha_M
)^2,
\ee
and the scalar equations for $\chi^\alpha$,
\be
\label{3-27}
\nabla_M (e^{\alpha_i \phi^i} F^{\alpha M} ) = \sum_{\beta-\gamma=-\alpha} N_{\beta,-\gamma}
e^{\gamma_i \phi^i} F^\beta_M F^{\gamma M}.
\ee

Next, we will proceed with finding an appropriate embedding of an $\frac{SL(2,\mathbb{R})}{U(1)}$
submanifold in the scalar cosets. The embedding proceeds by finding a subset $(h , e_+ )$ of the solvable Lie algebra generators (or suitable linear combinations) so that the closed set $(h,e_+,e_-)$, with $e_-$ being the negative-root generator corresponding to $e_+$, is normalized so as to satisfy the $\mathfrak{sl}(2,\mathbb{R})$ algebra. For the $SO(2,n)$ and $SO(3,n)$ cases, it is known that such a set is given by $(\alpha \cdot H , E_\alpha )$ where $\alpha$ can be any of the long roots; for the $SO(1,n)$ case, this requires a minor modification. The $\frac{SL(2,\mathbb{R})}{U(1)}$ submanifold will then be parameterized by the two fields that correspond to $h$ and $e_+$. For the three cases under consideration, the explicit embeddings are shown below.

\begin{itemize}

\item $D=9$. In the $D=9$ case, the desired embedding in $SO(1,n)/SO(n)$ is found by observing that the generators
\be
\label{3-28}
h = 2 H_{\epsilon_1} ,\qquad e_\pm = E_{\pm(\epsilon_1 + \epsilon_2)} + E_{\pm(\epsilon_1 -
\epsilon_2)},
\ee
satisfy the $\mathfrak{sl}(2,\mathbb{R})$ algebra
\be
\label{3-29}
[h_,e_\pm] = \pm 2 e_\pm ,\qquad
[e_+,e_-] = h.
\ee
The fields $\phi$ and $\chi$ parameterizing the $\frac{SL(2,\mathbb{R})}{U(1)}$ submanifold are the fields along $h$ and $e_+$ respectively, namely $\phi \equiv \frac{1}{2} \phi^1$ and $\chi \equiv \frac{1}{2} ( \chi^{\epsilon_1 + \epsilon_2} + \chi^{\epsilon_1 - \epsilon_2})$. In this case, it is readily shown that the scalar field equations are consistent with the configuration
\bea
\label{3-30}
\phi^i=0&& ~~\mbox{except}~~\phi(x) = \frac{1}{2} \phi^1(x) ~~~~~~ \nn\\
\chi^\alpha=0&&  ~~\mbox{except}~~ \chi(x) = \frac{1}{2} \Big{(} \chi^{\epsilon_1 + \epsilon_2} (x) + \chi^{\epsilon_1 - \epsilon_2}(x) \Big{)},
\eea
i.e., with a configuration where all fields except  $\phi$ and $\chi$ are zero. This can be seen by recalling that, in this particular case, all structure constants $C^\alpha_{\beta\gamma}$ vanish. It follows that (i) all axion field strengths are simply $F^\alpha_M = \partial_M \chi^\alpha$ and (ii) there are no nonzero $N_{\beta,-\gamma}$'s on the RHS of (\ref{3-27}) since that would require that $\alpha + \beta = \gamma$ i.e. that there exists a nonzero $C^\gamma_{\alpha\beta}$ for some $\gamma$. So, the RHS of (\ref{3-27}) is identically zero for all $\alpha$ and thus one is allowed to set any axion fields to zero.

\item $D=8$. In this case, the desired embedding in $\frac{SO(2,n) }{ SO(2) \times SO(n)}$ is identified by noticing that the generators
\be \label{3-31} h = H_{\epsilon_1} + H_{\epsilon_2} = H_{\epsilon_1 + \epsilon_2} ,\qquad e_\pm = E_{\pm(\epsilon_1 + \epsilon_2)},
\ee
satisfy the $\mathfrak{sl}(2,\mathbb{R})$ algebra. The fields parameterizing the $\frac{SL(2,\mathbb{R})}{U(1)}$ subspace are accordingly given by $\phi \equiv \frac{1}{2} (\phi^1 + \phi^2)$ and $\chi\ \equiv \chi^{\epsilon_1+\epsilon_2}$; the field strength of $\chi$ will be denoted as $F_M$.
Similarly to the $D=9$ case, to show that the configuration
\bea
\label{3-32}
\phi^i=0&&
~~\mbox{except}~~\phi(x) = \frac{1}{2}
\Big{(}\phi^1(x) + \phi^2(x)\Big{)} ~~~~~~ \nn\\
\chi^\alpha=0&&  ~~\mbox{except}~~\chi(x)= \chi^{\epsilon_1+\epsilon_2}(x)
\eea
where only $\phi$ and $\chi$ are nonzero is consistent with the equations of motion, we proceed as follows. Recalling that there are no no nonzero structure constants with lower index $\epsilon_1+\epsilon_2$, we immediately see that the only field strength containing $\chi$ is $F_M$ and has the form
\be
\label{3-33} F_M =
\partial_M \chi + \ldots
\ee
where the omitted terms are independent of $\chi$. It follows that, for our configuration, we
have $F_M = \partial_M \chi$ and $F^{\alpha \ne \epsilon_1+\epsilon_2} = 0$. Using these facts, we first consider
the equation of motion (\ref{3-26}) for the second dilaton $\phi' = \frac{1}{2} ( \phi_1 - \phi_2 )$. It is easily seen that the RHS of this equation vanishes,
\be
\label{3-34}
\frac{1}{2} \sum_{\alpha} ( \alpha_1 - \alpha_2 ) e^{\alpha_i \phi^i} ( F^\alpha_M )^2 = \frac{1}{2} [ ( \epsilon_1+\epsilon_2 )_1 - ( \epsilon_1+\epsilon_2 )_2 ] e^{2 \phi} ( F_M )^2 = 0, \ee and so
this equation is consistent with  $\phi'=0$.  Second, we consider the equation of motion (\ref{3-27}) for the axions $\chi^{\alpha \ne \epsilon_1+\epsilon_2}$. For our configuration, the LHS of this equation equals zero. The only possible case in which the RHS would be nonzero is the case where $\beta = \gamma = \epsilon_1+\epsilon_2$ so that the combination $(F_M)^2$ would  appear; however, this would require that $\alpha = -(\beta - \gamma) = 0$ which cannot be satisfied. Therefore, the configuration under consideration satisfies the scalar equations of motion.

\item $D=7$. In this last case, the embedding of $\frac{SL(2,\mathbb{R})}{U(1)}$ in $\frac{SO(3,n)}{ SO(3)
\times SO(n)}$ is again given by the generators $h$ and $e_\pm$ of (\ref{3-31}) and it is parametrized by the associated fields $\phi \equiv \frac{1}{2} (\phi^1 + \phi^2)$ and $\chi\ \equiv \chi^{\epsilon_1+\epsilon_2}$.

Again, the configuration (\ref{3-32}) is consistent with the equations of motion. For $\phi'$ and $\chi^{\alpha \ne \epsilon_1+\epsilon_2}$, the proof proceeds exactly as before. As for the extra dilaton
$\phi^3$ present in this case, the same reasoning leading to Eq. (\ref{3-35}) implies that the RHS of the corresponding equation of motion is given by
\be \label{3-35} \sum_{\alpha} \alpha_3
e^{\alpha_i \phi^i} ( F^\alpha_M )^2 = ( \epsilon_1+\epsilon_2 )_3
e^{2 \phi} ( F_M )^2 = 0,
\ee
and so it is again consistent to set $\phi^3=0$.

\end{itemize}

\subsection{The solution}
\label{sec-3-3}

As we have seen, the configurations (\ref{3-30}),(\ref{3-32}) in $D=9$ and  $D=8,7$, respectively are consistent with the scalar field equations.  It remains now to solve the Einstein equations (\ref{3-25}), as well as the equations (\ref{3-26},\ref{3-27}) for the remaining scalars $\phi,\chi$. By assembling $\phi$ and $\chi$ into the complex combination
\be
\label{3-36}
\tau = \tau_1 + \ii \tau_2 = \chi + \ii e^{-\phi},
\ee
the Einstein and scalar field equations may be written as
\be
\label{3-37} R_{MN} = \frac{1}{4 \tau_2^2} ( \partial_M \tau \partial_N \bar{\tau} + \partial_M \bar{\tau}
\partial_N \tau )
\ee
and
\be
\label{3-38}
\nabla^2\tau-\frac{2\d_M\tau\d^M\tau}{\bar{\tau}-\tau}=0,
\ee
respectively.  To solve these equations, we split the spacetime coordinates as $(x^\mu,y^m)$ where $x^\mu,~ \mu=0,\ldots,D-3$ parametrize a $(D\!-\!2)$--dimensional spacetime  and $y^m,~ m=1,2$ parametrize a two-dimensional surface. Writing the internal coordinates $y^m$ in the complex basis $(z,\bar{z})$, we use the ansatz
\be
\label{3-39} ds_D^2 = g_{\mu\nu} dx^\mu dx^\nu + e^{2 \Omega(z,\bar{z})} dz d\bar{z} ,\qquad \tau =
\tau(z,\bar{z})
\ee
where the scalars depend only on the internal coordinates. The scalar equation (\ref{3-38}) for this ansatz is written as
\be
\label{3-40} \d\bar{\d}\tau-\frac{2\d\tau\bar{\d}\tau}{\tau-\bar{\tau}}=0,
\ee
and is thus solved for any holomorphic (antiholomorphic) $\tau=\tau(z)~ \big{(}\tau(\bar{z})\big{)}$. Passing
to the Einstein equation, its $(\mu\nu)$ components are given by
\be
\label{3-41}
R_{\mu\nu} = 0,
\ee
which implies that we may take the $(D-2)$--dimensional spacetime to be Minkowski spacetime. As for the $(mn)$ components of the Einstein equation, they lead, for holomorphic $\tau$, to the equation
\be
\label{3-42}
- 2 \partial \bar{\partial} \Omega = \frac{\partial \tau \bar{\partial} \bar{\tau} }{4 \tau_2^2},
\ee
which is solved by
\be
\label{3-43}
\Omega = \frac{1}{2} \ln \tau_2 + f(z) + \bar{f}(\bar{z}),
\ee
where $f(z)$ can be any holomorphic function. Thus, the solution for the metric and the scalars reads
\be
\label{3-44}
ds_D^2 = \eta_{\mu\nu} dx^\mu dx^\nu + d\sigma^2\, , ~~~~d\sigma^2=\tau_2(z,\bar{z})|F(z)|^2 dz
d\bar{z} ,\qquad \tau=\tau(z),
\ee
where $F(z)=\exp\big{(}f(z)\big{)}$. Note that, for any holomorphic $\tau=\tau(z)$, there exists a corresponding 2D metric $d\sigma^2$ (\ref{3-44}), which may be singular or non-singular.

\subsubsection{Singular solutions}

A homomorphic $\tau$, which leads to a singular solution is \cite{Kehagias:2004fb}
\be
\label{3-45}
\tau= \ii \frac{R^b+iz^b}{R^b-iz^b} .
\ee
This field gives rise  to the 2D metric
\be
\label{3-46}
d\sigma^2 = \left(1-\Big{|}\frac{z}{R}\Big{|}^{2b}\right)dz d\bar{z} \,
\ee
after choosing $F(z)=1$ for the metric to be regular around $z=0$. This solution, for $b=1$, reduces to the tear-drop \cite{Gell-Mann1},\cite{Gell-Mann2} and it is singular at $|z|=R$ where it has a naked singularity. However, the singularity is harmless since it does not lead to any physically unacceptable situation. It can be proven for example that energy, momentum and angular momentum are conserved \cite{Gell-Mann1},\cite{Gell-Mann2}, whereas, it may be relevant for the solution of the cosmological constant problem \cite{Kehagias:2004fb}. Moreover, the volume of the singular transverse space is finite, which leads to finite 4D Planck constant and thus to conventional 4D gravitational interactions.

\subsubsection{Non-singular solutions}

There are also non-singular solutions to the equations (\ref{3-40},\ref{3-42}). For the construction of these solutions, we recall that the field equations (\ref{3-37},\ref{3-38}) are invariant under the $SL(2,\mathbb{R})$ transformation
\be
\label{3-47}
\tau \to \frac{a \tau+b}{c\tau+d}\, , ~~~~~ad-cb=1\, ,
\ee
for real $a,b,c,d$. In fact, the symmetry is reduced to the modular group $SL(2,\mathbb{Z})$,  when non-perturbative effects are taken into account. To proceed, we note that the energy per unit $(D\!-\!2)$--volume is
\be
\label{3-48}
E= - \frac{\ii}{2}\int d^2 z \partial\bar{\partial}\ln \tau_2 \, .
\ee
In order to find finite energy solutions one has to restrict $\tau$ to the fundamental domain of $SL(2,\mathbb{Z})$~\cite{GV}. Then, $\tau$ has discontinuous jumps done by the $SL(2,\mathbb{Z})$ transformations $\tau\rightarrow \tau+1$ as we go around the singularities at $z=z_i$.  These jumps and the requirement of holomorphicity imply that, near the location of the singularities, we must have
\be
\label{3-49}
\tau \simeq \frac{1}{2\pi \ii}\ln (z-z_i) \, .
\ee
The energy in this case is indeed finite and it turns out to be proportional to the volume of the fundamental domain ${\cal F}_1$,
\be
\label{3-50}
E=\frac{\pi}{6}n \, ,
\ee
where $n$ is the number of  times the z-plane covers ${\cal F}_1$.

Since the fundamental domain  of $SL(2,\mathbb{Z})$ is mapped to the complex sphere in the $j$--plane through the modular $j$--function, we may express the solution for $\tau$ as the pull-back of $j(\tau)$. Thus we may write
\be
\label{3-51}
j(\tau)=\frac{P(z)}{Q(z)} \, ,
\ee
where $P(z),Q(z)$ are polynomials of degree $p$ and $q$, respectively. If $p\leq q$, $j$ approaches a constant value as $|z|\rightarrow \infty$ and $n=q$ in this case. There exist $q$ points at which $Q(z)$ has
zeroes and these points are singular. As has been shown in~\cite{Nair:2004yu}, there are singularities
at the zeroes of $P(z)$ as well, which an be avoided however, by choosing $P(z)=\mbox{const}.$.

Recalling now that $SL(2,\mathbb{Z})$ is generated by
\be
\label{3-52}
\tau \to -\frac{1}{\tau} ,\qquad \tau\to \tau+1\, ,
\ee
the metric (\ref{3-44}) is clearly not modular invariant. However, we may use the freedom to choose the holomorphic function $F(z)$ to make the metric non-degenerate as well as modular invariant. These
two conditions specify $F(z)$ to be
\be
\label{3-53}
F(z)=\eta(\tau)^2 \prod_{i=1}^{n}(z-z_i)^{-1/12} \, ,
\ee
where $\eta(\tau)=q^{1/24}\prod_{r>0}(1-q^r)$ is Dedekind's $\eta$--function ($q=e^{2\pi i\tau}$). Then, the  metric turns out to be
\be
\label{3-54}
ds_D^2 = \eta_{\mu\nu} dx^\mu dx^\nu+\tau_2\eta(\tau)^2\bar{\eta}(\bar{\tau}) \Big{|}
\prod_{i=1}^{n}(z-z_i)^{-1/12}\Big{|}^2 dz d\bar{z}\, .
\ee
The asymptotic form of (\ref{3-54})  is
\be
\label{3-55}
ds_D^2\sim \eta_{\mu\nu} dx^\mu dx^\nu+(z\bar{z})^{-n/12}dzd\bar{z} \, ,
\ee
and one recognizes  a deficit angle $\delta=\pi n/6$. With $n=12$ strings the deficit angle becomes $\delta=2\pi$ and the transverse space is asymptotically a cylinder while $n=24$ strings produce a deficit angle
$\delta=4\pi$ and the transverse space is a compact $S^2$. As a result, the $\tau$--field configurations  defined implicitly by
\be
\label{3-56}
j(\tau(z))=\prod_{i=1}^{24}\frac{1}{z-z_i}
\ee
compactifies the D-dimensional space-time to $M^{D-2}\times S^2$ with metric given in eq.(\ref{3-54}) for $n=24$.

We should mention here that there are some special cases which may further compactify $M^D$. For example, one may consider two  vector multiplets coupled to the gravity multiplet in $D=8$ supergravity. In this case, the scalar manifold is $\frac{SO(2,2)}{SO(2)\times SO(2)}$, which is actually $\frac{SL(2,\mathbb{R})}{U(1)}\times
\frac{SL(2,\mathbb{R})}{U(1)}$. Then, in this case, compactification to four dimensions may be achieved and the $D=8$ vacuum is of the form $M^4\times S^2\times S^2$ with metric
\bea
\label{3-57}
ds_D^2 &=& \eta_{\mu\nu} dx^\mu dx^\nu \\ &&+\tau_2\eta(\tau)^2\bar{\eta}(\bar{\tau}) \Big{|}
\prod_{i=1}^{n}(z-z_i)^{-1/12}\Big{|}^2 dz d\bar{z} + \sigma_2\eta(\sigma)^2\bar{\eta}(\bar{\sigma})\Big{|}
\prod_{i=1}^{n}(w-w_i)^{-1/12}\Big{|}^2 dw d\bar{w} \, .\nonumber
\eea
where $\tau,\sigma$ parametrize the two $\frac{SL(2,\mathbb{R})}{U(1)}$ factors of the scalar manifold and $z,w$are the complex coordinates on the
transverse 4D space.

\section{Supersymmetry}
\label{sec-4}

The last issue we intend to address is how many of the supersymmetries of the original theories are preserved in our scalar-induced
compactifications.
In what follows, we will show that the compactification of the $D=9,8,7$ theories preserve one half of the initial supersymmetries, leading to
effective theories with $D=7$, $\MN=2$, $D=6$, $\MN=1$ and $D=5$, $\MN=2$ supersymmetry respectively.

\subsection{$D=9$}
\label{sec-4-1}

In the $D=9$ case, all we need to do is consider the gravitino
variation which, in a background where all fields except gravity
and $\tau$ are zero, is given by \be \label{4-1} \delta \psi_M =
\nabla_M \epsilon = ( \partial_M + \omega_{M} ) \epsilon, \ee
where\footnote{Here and in what follows, the flat-space gamma
matrices are denoted as $\gamma^{A}$ and the curved-space gamma
matrices are denoted as $\Gamma^M = e^M_A \gamma^A$.} $\omega_M
\equiv \frac{1}{4} \omega_{M AB} \gamma^{AB}$. Due to the specific
form of the metric in (\ref{3-44}), the only non-trivial
spin-connection terms are $\omega_m$. Using conformal flatness of
the internal metric, we find \be \label{4-2} \omega_m =
\frac{1}{2} \gamma_m^{\phantom{m} n} \partial_n \Omega \, , \ee
and, thus, for
\be
\gamma^{78}\epsilon=-\ii \epsilon
\ee
 %spinors $\epsilon_{\pm}$ as \be \label{4-3}
%\gamma_{\bar{z}}\epsilon_+=0 ,\qquad \gamma_z\epsilon_-=0, \ee
we get for the gravitino shifts  \be
\label{4-4} \delta \psi_z = \left( \partial +
\frac{1}{2}\partial\Omega \right) \epsilon ,\qquad \delta
\psi_{\bar{z}} = \left( \bar{\partial}  -  \frac{1}{2}
\bar{\partial} \Omega\right) \epsilon. \ee It easy to see that
the vanishing of the gravitino shift $\delta \psi_{\bar{z}}$
specify $\epsilon_+$ to be \be \label{4-5}
\epsilon_+=e^{\frac{1}{2}\Omega}\epsilon_0 \ee where $\epsilon_0$
is a constant spinor. Then, $\delta \psi_z$ is then non-vanishing
and the solution preserves half of the supersymmetries. It can
easily be checked that $\delta \psi_z$ is normalizable as it
should.

\subsection{$D=8$}
\label{sec-4-2}

To check supersymmetry of our compactification in the $D=8$ case, we have to consider the non-trivial
supersymmetry variations of the fields. These are the variation of the gravitino, which now has
the form
\be \label{4-6} \delta \psi_M = D_M \epsilon = \left(  \partial_M + \omega_{M} - \frac{\ii}{2} Q_M \right) \epsilon,
\ee
and the variation of the gauginos, given by
\be
\label{4-7} \delta
\lambda_{\bar{a}} = - \frac{\ii}{2} \Gamma^M \hat{P}_{M \bar{a}} \epsilon,
\ee
where $Q_M$ and $\hat{P}_{M \bar{a}}$ are given in (\ref{3-16}). To check whether the above equations are satisfied in our background, we have to compute $\omega_m$, $Q_m$ and $\hat{P}_{m a}$ (since all $\mu$ components vanish). Regarding the spin connection, we find
\be
\label{4-8}
\omega_z = \frac{\ii}{2} \gamma^{67} \partial \Omega ,\qquad \omega_{\bar{z}} = - \frac{\ii}{2} \gamma^{67}\bar{\partial} \Omega .
\ee
Regarding $Q_m$ and $\hat{P}_{m \bar{a}}$, they may be read off from the Maurer-Cartan form in the standard $SO(2,n)$ basis according to (\ref{3-15}). To this end, we take for definiteness the two non-compact Cartan
generators of $SO(2,n)$ as $H_{\epsilon_1} = T_{13}$ and $H_{\epsilon_2} = T_{24}$, where $(T_{IJ})_{KL} \equiv \delta_{IK} \delta_{JL} + \delta_{IL} \delta_{JK}$. In this basis, the generators $H_{\epsilon_1+\epsilon_2}$ and $E_{\epsilon_1 + \epsilon_2}$ are given by \cite{Helgason},\cite{Castellani:1998nz}
\be
\label{4-9}
H_{\epsilon_1+\epsilon_2} = \left(
\begin{array}{ccccc}
0 & 0 & 1 & 0 &   \cdots \\
0 & 0 & 0 & 1 &   \cdots \\
1 & 0 & 0 & 0 & \cdots\\
0 & 1 & 0 & 0 &   \cdots \\
\vdots & \vdots & \vdots & \vdots & \ddots \\
\end{array} \right) , \qquad
E_{\epsilon_1 + \epsilon_2} = \frac{1}{2} \left(
\begin{array}{ccccc}
0  & -1 & 0  & 1 & \cdots \\
1  &  0 & -1 & 0  & \cdots \\
0  & -1 & 0  & 1 & \cdots\\
1  &  0 & -1 & 0  & \cdots \\
\vdots & \vdots & \vdots & \vdots & \ddots \\
\end{array} \right).
\ee
Inserting the above expressions into (\ref{2-9}), now given by $\partial_M L L^{-1}
= \frac{1}{2} \partial_M \phi H_{\epsilon_1 + \epsilon_2} + e^\phi \partial_M \chi E_{\epsilon_1 + \epsilon_2}$,
we find the Maurer-Cartan form
\be \label{4-10}
\partial_M L L^{-1} =
\frac{1}{2} \left( \begin{array}{cccccc}
0                        & - e^\phi \partial_M \chi     & \partial_M \phi      & e^\phi \partial_M \chi & 0 & \cdots \\
e^\phi \partial_M \chi   & 0                            & - e^\phi \partial_M \chi & \partial_M \phi         & 0 & \cdots \\
\partial_M \phi        & - e^\phi \partial_M \chi   & 0                        & e^\phi \partial_M \chi  & 0 & \cdots\\
e^\phi \partial_M \chi & \partial_M \phi            & - e^\phi \partial_M \chi & 0                         & 0 & \cdots \\
0 & 0 & 0 & 0 & 0 & \cdots \\
\vdots & \vdots & \vdots & \vdots & \vdots & \ddots \\
\end{array} \right).
\ee
From this expression, we read off the $SO(2)$ connection
\be
\label{4-11}
Q_m = - \frac{1}{2} e^\phi \partial_m \chi = -\frac{\partial_m \tau_1}{2 \tau_2},
\ee
and the vielbein
\bea
\label{4-12}
\hat{P}_{m 1} =  - \frac{\ii}{2 \tau_2} ( \gamma_{(9)} \partial_m \tau_1 - \ii \partial_m \tau_2 )\, , ~~~~
\hat{P}_{m 2} =
 \frac{1}{2 \tau_2} ( \partial_m \tau_1 - \ii \gamma_{(9)} \partial_m \tau_2 ).
\eea

Using the above expressions, we can determine the supersymmetry of the background. Starting
from the gravitino variation (\ref{4-6}), we write its $z$ and $\bar{z}$ components as \
\bea
\label{4-13}
\delta \psi_z = \left( \partial \!+\! \frac{\ii}{4} \frac{\partial \tau_1 + \gamma^{67} \partial \tau_2}{\tau_2}\! +\!
\frac{\ii}{2} \gamma^{67} \partial f \right)\! \epsilon ,~
\delta \psi_{\bar{z}} = \left( \bar{\partial}\! +
\! \frac{\ii}{4} \frac{\bar{\partial} \tau_1 \!-\! \gamma^{67}
\bar{\partial} \tau_2}{\tau_2} - \frac{\ii}{2} \gamma^{67} \bar{\partial} \bar{f} \right)\! \epsilon
\eea
Unlike the previous case, both components of the gravitino variation can be made to vanish for nontrivial $\epsilon$. Indeed, we immediately see that, if $\epsilon$ is subject to the condition
\be \label{4-14}
\gamma^{67} \epsilon = - \ii \epsilon,
\ee
then the $SO(2)$ connection cancels the $\tau_2$--dependent part of the spin connection for holomorphic $\tau$. Then, the variation (\ref{4-13}) vanishes if $\epsilon$ is given by
\be
\epsilon = e^{- \ii f_2(z,\bar{z})} \epsilon_0,
\ee
where $\epsilon_0$ is a constant spinor subject to (\ref{4-14}) and $f_2(z,\bar{z}) = \Imag f(z)$. Turning to the gaugino variation (\ref{4-7}), we note that the supersymmetry spinor $\epsilon$ satisfies
\be
\label{4-15} \gamma_{(9)} \epsilon = \epsilon.
\ee
Then, noticing that (\ref{4-14}) implies that $(\gamma^6 - \ii \gamma^7) \epsilon = 0$, we easily find that the gaugino variation vanishes as well,
\bea \label{4-16}
\delta \lambda_{\bar{1}} = - \frac{1}{4 \tau_2} \Gamma^m \partial_m \bar{\tau} \epsilon
=  0, ~~~~
\delta \lambda_{\bar{2}} = - \frac{\ii}{4 \tau_2} \Gamma^m \partial_m \bar{\tau} \epsilon
= 0.
\eea
The constraint imposed by (\ref{4-14}) on the spinor $\epsilon$ projects out half its components and amounts to
a chirality projection. To see this, we note that the 8D and 6D chirality operators $\gamma_{(9)}$ and $\gamma_{(7)}$ are related by
\be
\label{4-17}
\gamma_{(9)} = -\ii \gamma_{(7)} \gamma^{67},
\ee
so that Eq. (\ref{4-15}) is equivalent to a chirality projection in 6D,
\be
\label{4-18}
\gamma_{(7)} \epsilon = - \epsilon.
\ee
Therefore, the background (\ref{3-44}) preserves half the supersymmetries of the original theory, leading to a
chiral 6D effective theory with $\MN=1$ supersymmetry.

\subsection{$D=7$}
\label{sec-4-3}

Let us finally check the supersymmetry of our compactification in the $D=7$ case. Here, the non-trivial
supersymmetry variations of the fields are given by the gravitino variation
\be
\label{4-19} \delta \psi_{M\bar{i}} = D_M \epsilon_{\bar{i}} = ( \partial_M +
\omega_{M} ) \epsilon_{\bar{i}} + \frac{1}{2} Q_{M
\bar{i}}^{\phantom{M \bar{i}} \bar{j}} \epsilon_{\bar{j}},
\ee
and the gaugino variation
\be \label{4-20}
\delta \lambda_{\bar{a} \bar{i}} = \ii \sqrt{2} \Gamma^M P_{M \bar{a} \bar{i}}^{\phantom{M
\bar{a}\bar{i}} \bar{j} } \epsilon_{\bar{j}}.
\ee
Proceeding as before, we find the spin connection $\omega_z = \frac{\ii}{2} \gamma^{56}\partial \Omega$.
%\be
%\label{4-21}
%\omega_z = \frac{\ii}{2} \gamma^{56} \left( \frac{\partial \tau_2}{2 \tau_2} + \partial f \right) ,\qquad \omega_{\bar{z}} = - \frac{\ii}{2} \gamma^{56} \left(
%\frac{\bar{\partial} \tau_2}{2 \tau_2} + \bar{\partial} \bar{f} \right) .
%\ee
Passing employing a basis where $H_{\epsilon_1} = T_{24}$ and $H_{\epsilon_2} = T_{35}$, we find the Maurer-Cartan form
\be
\label{4-22}
\partial_M L L^{-1} =
\frac{1}{2} \left( \begin{array}{ccccccc}
0 & 0 & 0 & 0 & 0 & 0 & \cdots \\
0 & 0                        & - e^\phi \partial_M \chi     & \partial_M \phi      & e^\phi \partial_M \chi & 0 & \cdots \\
0 & e^\phi \partial_M \chi   & 0                            & - e^\phi \partial_M \chi & \partial_M \phi         & 0 & \cdots \\
0 & \partial_M \phi        & - e^\phi \partial_M \chi   & 0                        & e^\phi \partial_M \chi  & 0 & \cdots\\
0 & e^\phi \partial_M \chi & \partial_M \phi            & - e^\phi \partial_M \chi & 0                         & 0 & \cdots \\
0 & 0 & 0 & 0 & 0 & 0 & \cdots \\
\vdots & \vdots & \vdots & \vdots & \vdots & \vdots & \ddots \\
\end{array} \right).
\ee
From this, we can immediately read off the $SO(3)$ connection and the vielbein in the forms $\widetilde{Q}_{m
\bar{X}}^{\phantom{m \bar{X}} \bar{Y}}$ and $\widetilde{P}_{m \bar{a}}^{\phantom{m \bar{a}}
\bar{X}}$ that involve the \emph{triplet} index $\bar{X}=1,2,3$ for $SO(3)$ (the use of this
index is emphasized by the tildes). We find
\be
\label{4-23}
\widetilde{Q}_{m \bar{2}}^{\phantom{m \bar{2}} \bar{3}} = -
\widetilde{Q}_{m \bar{3}}^{\phantom{m \bar{3}} \bar{2}} = -
\frac{\partial_m \tau_1}{2 \tau_2} \ee and \be \label{4-24}
\widetilde{P}_{m \bar{1}}^{\phantom{m \bar{1}} \bar{2}} =
\widetilde{P}_{m \bar{2}}^{\phantom{m \bar{2}} \bar{3}} = -
\frac{\partial_m \tau_2}{2 \tau_2} ,\qquad \widetilde{P}_{m
\bar{1}}^{\phantom{m \bar{1}} \bar{3}} = - \widetilde{P}_{m
\bar{2}}^{\phantom{m \bar{2}} \bar{2}} = - \frac{\partial_m
\tau_1}{2 \tau_2}
\ee
To insert these expressions in the supersymmetry transformations (\ref{4-19}) and (\ref{4-20}), we
have to switch from the triplet notation to the doublet notation. For this, we use the transformations
\be
\label{4-25}
Q_{m \bar{i}}^{\phantom{m \bar{i}} \bar{j}} = \frac{\ii}{2} \epsilon_{\bar{X}\bar{Y}\bar{Z}} (\sigma^{\bar{X}})_{\bar{i}}^{\phantom{\bar{i}} \bar{j}} \widetilde{Q}_m^{\phantom{m} \bar{Y}\bar{Z}} ,\qquad P_{m \bar{a} \bar{i}}^{\phantom{m \bar{a}\bar{i}} \bar{j} } = \frac{1}{\sqrt 2} \widetilde{P}_{m \bar{a}}^{\phantom{m \bar{a}}
\bar{X}} (\sigma^{\bar{X}})_{\bar{i}}^{\phantom{\bar{i}} \bar{j}},
\ee
which yield
\be
\label{4-26}
Q_{m \bar{1}}^{\phantom{m \bar{1}} \bar{2}} = Q_{m \bar{2}}^{\phantom{m \bar{2}} \bar{1}} =
\ii \frac{\partial_m \tau_1}{2 \tau_2} ,
\ee
and
\bea
\label{4-27}
&& P_{m \bar{1} \bar{1}}^{\phantom{m \bar{1}\bar{1}} \bar{1} } = -
P_{m \bar{1} \bar{2}}^{\phantom{m \bar{1}\bar{2}} \bar{2} } = -
\frac{\partial_m \tau_1}{2 \sqrt{2} \tau_2} ,\qquad P_{m \bar{1}
\bar{1}}^{\phantom{m \bar{1}\bar{1}} \bar{2} } = - P_{m \bar{1}
\bar{2}}^{\phantom{m \bar{1}\bar{2}} \bar{1} } = \ii \frac{\partial_m \tau_2}{2 \sqrt{2} \tau_2}, \nn\\
&& P_{m \bar{2} \bar{1}}^{\phantom{m \bar{2}\bar{1}} \bar{1} } = - P_{m \bar{2} \bar{2}}^{\phantom{m \bar{2}\bar{2}}
\bar{2} } = - \frac{\partial_m \tau_2}{2 \sqrt{2} \tau_2} ,\qquad P_{m \bar{2} \bar{1}}^{\phantom{m \bar{2}\bar{1}} \bar{2} }
= - P_{m \bar{2} \bar{2}}^{\phantom{m \bar{2}\bar{2}} \bar{1} } = - \ii \frac{\partial_m
\tau_1}{2 \sqrt{2} \tau_2}.
\eea

Using the above expressions, we can determine the supersymmetry of the background.
Indeed, considering the gravitino shifts (\ref{4-19}) we find that, $\delta \psi_{M\bar{i}} =0$
%The gravitino variation turns out to be
%\bea
%\label{4-28}
%\delta \psi_{z\bar{1}} &=& \left( \partial + \frac{\ii}{2} \gamma^{56}
%\partial f \right) \epsilon_{\bar{1}} + \frac{\ii}{4 \tau_2} ( \partial
%\tau_1 \epsilon_{\bar{2}} + \gamma^{56} \partial \tau_2 \epsilon_{\bar{1}} ) , \nn\\
%%\delta \psi_{z\bar{2}} &=& \left( \partial + \frac{\ii}{2} \gamma^{56}
%\partial f \right) \epsilon_{\bar{2}} + \frac{\ii}{4 \tau_2} ( \partial
%\tau_1 \epsilon_{\bar{1}} + \gamma^{56} \partial \tau_2 \epsilon_{\bar{2}} ) , \nn\\
%%\delta \psi_{\bar{z} \bar{1}} &=& \left( \bar{\partial} - \frac{\ii}{2}
%\gamma^{56} \bar{\partial} \bar{f} \right) \epsilon_{\bar{1}} +
%\frac{\ii}{4 \tau_2} ( \bar{\partial} \tau_1 \epsilon_{\bar{2}} -
%\gamma^{56} \bar{\partial} \tau_2 \epsilon_{\bar{1}} ). \nn\\
%%\delta \psi_{\bar{z} \bar{2}} &=& \left( \bar{\partial} - \frac{\ii}{2}
%\gamma^{56} \bar{\partial} \bar{f} \right) \epsilon_{\bar{2}} + \frac{\ii}{4 \tau_2} ( \bar{\partial} \tau_1 \epsilon_{\bar{1}} - \gamma^{56} \bar{\partial} \tau_2 \epsilon_{\bar{2}} ).
%\eea
if the spinors $\epsilon_{\bar{1}}$ and $\epsilon_{\bar{2}}$ are subject to the condition
\be
\label{4-29}
\gamma^{56} \epsilon_{\bar{1}} = - \ii \epsilon_{\bar{2}}.
\ee
Then the $SO(3)$ connection cancels the $\tau_2$--dependent part of the spin connection
for holomorphic $\tau$.
The resulting equations form a system of coupled differential equations whose solution is concisely written as
\be
\epsilon_{\bar{1}} = e^{\gamma^{56} f_2(z,\bar{z})} \epsilon_{0,\bar{1}} ,\qquad \epsilon_{\bar{2}} = e^{\gamma^{56} f_2(z,\bar{z})} \epsilon_{0,\bar{2}},
\ee
where $\epsilon_{0,\bar{1}}$ and $\epsilon_{0,\bar{2}}$ are constant spinors satisfying (\ref{4-29}).
Then, it can easily  be shown that the gaugino shifts vanish as well.
% Turning to the gaugino variation, it is given by
%\bea
%\label{4-30}
%&& \!\!\!\!\!\!\!\! \delta \lambda_{\bar{1} \bar{1}} = - \frac{\ii}{2 \tau_2} \Gamma^m ( \partial_m \tau_1
%\epsilon_{\bar{1}} - \ii \partial_m \tau_2 \epsilon_{\bar{2}} ) ,\qquad \delta \lambda_{\bar{1} \bar{2}} =
% \frac{\ii}{2 \tau_2} \Gamma^m ( \partial_m \tau_1 \epsilon_{\bar{2}} - \ii \partial_m \tau_2 \epsilon_{\bar{1}} ), \nn\\
%&& \!\!\!\!\!\!\!\! \delta \lambda_{\bar{2} \bar{1}} = \frac{1}{2 \tau_2} \Gamma^m ( \partial_m \tau_1 \epsilon_{\bar{2}}
% - \ii \partial_m \tau_2 \epsilon_{\bar{1}} ) ,\qquad \delta \lambda_{\bar{2} \bar{2}} = -
%\frac{1}{2 \tau_2} \Gamma^m ( \partial_m \tau_1 \epsilon_{\bar{1}} - \ii \partial_m \tau_2 \epsilon_{\bar{2}} ).
%\eea
Noting that the condition (\ref{4-29}) implies that $\gamma^6 \epsilon_{\bar{1},\bar{2}} = - \ii \gamma^5
\epsilon_{\bar{2},\bar{1}}$ and using holomorphicity of $\tau$, one sees that this variation vanishes as well.
For example, for $\delta \lambda_{\bar{1} \bar{1}}$, we find
\bea
\label{4-31}
\delta \lambda_{\bar{1} \bar{1}} &=& - \frac{\ii}{2 \tau_2^{3/2}} \left[ \gamma^5 ( \partial_5 \tau_1
\epsilon_{\bar{1}} - \ii \partial_5 \tau_2 \epsilon_{\bar{2}} ) + \gamma^6 ( \partial_6 \tau_1
\epsilon_{\bar{1}} - \ii \partial_6 \tau_2 \epsilon_{\bar{2}} ) \right] \nn\\
&=& - \frac{\ii}{2 \tau_2^{3/2}} \gamma^5 \left[ ( \partial_5 \tau_1 - \partial_6 \tau_2 )
\epsilon_{\bar{1}} - \ii ( \partial_6 \tau_1 + \partial_5 \tau_2 ) \epsilon_{\bar{2}}
\right] \nn\\
&=& - \frac{\ii}{2 \tau_2^{3/2}} \gamma^5 \left[ ( \bar{\partial} \tau + \partial \bar{\tau} )
\epsilon_{\bar{1}} - ( \bar{\partial} \tau - \partial \bar{\tau} ) \epsilon_{\bar{2}}
\right] = 0,
\eea
and similarly for the other gauginos.

The condition (\ref{4-29}), correlating the two symplectic-Majorana 7D supersymmetry spinors, again halves their degrees of freedom. So, the background (\ref{3-44}) preserves half the supersymmetries of the original theory, leading to a 5D effective theory with $\MN=2$ supersymmetry.

\section{Conclusions}

In this work, we have shown how scalars may trigger compactification of $D=9,8$ and $D=7$ supergravities. This has previously be shown to work for the tear-drop solution~\cite{Gell-Mann1},\cite{Gell-Mann2} in 10D type IIB supegravity. In this case, the complex scalar of type IIB theory, which parametrize $\frac{SL(2,\mathbb{R})}{U(1)}$ is used to curl up two of the space-time coordinates leading to an internal space diffeomorphic to the scalar manifold. There is a naked singularity in this construction, which however
 is harmless as it does not lead to any violation of conservation laws. In other words, although the presence of the singularity is annoying, there is no any leakage of energy, momentum, or angular momentum through it. Moreover, the volume of the transverse external space is finite leading to a finite 4D Planck mass and, consequently, to conventional 4D gravity.

Here, we studied a similar possible mechanism for the case of higher-dimensional supergravities. The compactification mechanism we were after is triggered by the many scalars which exist in higher-dimensional supergravities. We have considered $D=9,8,7$ minimal supergravities coupled to vector multiplets. The vector multiplets contain scalars which, together with the scalars of the gravity multiplet, form scalar manifolds of the general form $\frac{SO(10\!-\!D,n)}{SO(10\!-\!D)\times SO(n)}$ in the presence of $n$ vector multiplets. We have shown here that there are solutions to the supergravity field equations where all but two of the scalars are non zero. These two non-trivial scalars parametrize the $\frac{SL(2,\mathbb{R})}{U(1)}$ coset. We then presented solutions which are either singular, like the tear-drop prototype~\cite{Gell-Mann1},\cite{Gell-Mann2},
or non-singular like the stringy-cosmic string~\cite{GV}. The latter compactify spacetime into $M^{D\!-\!2}\times S^2$, where for the particular case of $D=8$ with two vector multiplets, compactification to $M^{4}\times S^2\times S^2$ may be achieved. Finally, we have shown that these compactifications are supersymmetric.

\vskip .3in
\noindent
{\bf Acknowledgements}\\
We would like to thank S. D. Avramis for helpful discussions. This work is co-funded by the European Social Fund (75\%) and National Resources (25\%) - (EPEAEK II) - PYTHAGORAS.


\begin{thebibliography}{999}

\bibitem{Gell-Mann1}
  M.~Gell-Mann and B.~Zwiebach,
  ``Curling Up Two Spatial Dimensions With SU(1,1)/U(1),''
  Phys.\ Lett.\ B {\bf 147}, 111 (1984).
  %%CITATION = PHLTA,B147,111;%%

\bibitem{Gell-Mann2}
M.~Gell-Mann and B.~Zwiebach,
 ``Dimensional Reduction Of Space-Time Induced By Nonlinear Scalar Dynamics And
   Noncompact Extra Dimensions,''
   Nucl.\ Phys.\ B {\bf 260}, 569 (1985).

\bibitem{Nicolai}
  H.~Nicolai and C.~Wetterich,
  ``On The Spectrum Of Kaluza-Klein Theories With Noncompact Internal Spaces,''
  Phys.\ Lett.\ B {\bf 150}, 347 (1985).
  %%CITATION = PHLTA,B150,347;%%

\bibitem{GV}
  B.~R.~Greene, A.~D.~Shapere, C.~Vafa and S.~T.~Yau,
  ``Stringy Cosmic Strings And Noncompact Calabi-Yau Manifolds,''
  Nucl.\ Phys.\ B {\bf 337}, 1 (1990).
  %%CITATION = NUPHA,B337,1;%%

\bibitem{Gilmore}
  R.~Gilmore
  ``Lie Groups, Lie Algebras and some of their Applications,''
  Wiley, New York, (1974)

\bibitem{Helgason}
  S.~Helgason
  ``Differential Geometry, Lie Groups and Symmetric Spaces,''
  Academic Press, New York (1978).

%\cite{Andrianopoli:1996bq}
\bibitem{Andrianopoli:1996bq}
  L.~Andrianopoli, R.~D'Auria, S.~Ferrara, P.~Fre and M.~Trigiante,
  ``R-R scalars, U-duality and solvable Lie algebras,''
  Nucl.\ Phys.\ B {\bf 496}, 617 (1997)
  [arXiv:hep-th/9611014].
  %%CITATION = HEP-TH 9611014;%%

%\cite{Andrianopoli:1996zg}
\bibitem{Andrianopoli:1996zg}
  L.~Andrianopoli, R.~D'Auria, S.~Ferrara, P.~Fre, R.~Minasian and M.~Trigiante,
  ``Solvable Lie algebras in type IIA, type IIB and M theories,''
  Nucl.\ Phys.\ B {\bf 493}, 249 (1997)
  [arXiv:hep-th/9612202].
  %%CITATION = HEP-TH 9612202;%%

\bibitem{Arcioni:1998mn}
  G.~Arcioni, A.~Ceresole, F.~Cordaro, R.~D'Auria, P.~Fre, L.~Gualtieri and M.~Trigiante,
  ``N = 8 BPS black holes with 1/2 or 1/4 supersymmetry and solvable Lie
  algebra decompositions,''
  Nucl.\ Phys.\ B {\bf 542}, 273 (1999)
  [arXiv:hep-th/9807136].
  %%CITATION = HEP-TH 9807136;%%

\bibitem{Castellani:1998nz}
  L.~Castellani, A.~Ceresole, R.~D'Auria, S.~Ferrara, P.~Fre and M.~Trigiante,
  ``G/H M-branes and AdS(p+2) geometries,''
  Nucl.\ Phys.\ B {\bf 527}, 142 (1998)
  [arXiv:hep-th/9803039].
  %%CITATION = HEP-TH 9803039;%%

\bibitem{Lu:1998xt}
  H.~Lu, C.~N.~Pope and K.~S.~Stelle,
  ``M-theory/heterotic duality: A Kaluza-Klein perspective,''
  Nucl.\ Phys.\ B {\bf 548}, 87 (1999)
  [arXiv:hep-th/9810159].
  %%CITATION = HEP-TH 9810159;%%

\bibitem{Keurentjes:2002xc}
  A.~Keurentjes,
  ``The group theory of oxidation,''
  Nucl.\ Phys.\ B {\bf 658}, 303 (2003)
  [arXiv:hep-th/0210178].
  %%CITATION = HEP-TH 0210178;%%

%\cite{Keurentjes:2002rc}
\bibitem{Keurentjes:2002rc}
  A.~Keurentjes,
  ``The group theory of oxidation. II: Cosets of non-split groups,''
  Nucl.\ Phys.\ B {\bf 658}, 348 (2003)
  [arXiv:hep-th/0212024].
  %%CITATION = HEP-TH 0212024;%%

%\cite{Yilmaz:2003as}
\bibitem{Yilmaz:2003as}
  N.~T.~Yilmaz,
  ``Dualisation of the general scalar coset in supergravity theories,''
  Nucl.\ Phys.\ B {\bf 664}, 357 (2003)
  [arXiv:hep-th/0301236].
  %%CITATION = HEP-TH 0301236;%%

%\cite{Dereli:2004uh}
\bibitem{Dereli:2004uh}
  T.~Dereli and N.~T.~Yilmaz,
  ``Dualisation of the Salam-Sezgin D = 8 supergravity,''
  Nucl.\ Phys.\ B {\bf 691}, 223 (2004)
  [arXiv:hep-th/0407004].
  %%CITATION = HEP-TH 0407004;%%

%\cite{Tanii:1998px}
\bibitem{Tanii:1998px}
  Y.~Tanii,
  ``Introduction to supergravities in diverse dimensions,''
  arXiv:hep-th/9802138.
  %%CITATION = HEP-TH 9802138;%%

%\cite{Gates:1984kr}
\bibitem{Gates:1984kr}
  S.~J.~Gates, H.~Nishino and E.~Sezgin,
  ``Supergravity In D = 9 And Its Coupling To Noncompact Sigma Model,''
  Class.\ Quant.\ Grav.\  {\bf 3}, 21 (1986).
  %%CITATION = CQGRD,3,21;%%

%\cite{Salam:1985ns}
\bibitem{Salam:1985ns}
  A.~Salam and E.~Sezgin,
  ``D = 8 Supergravity: Matter Couplings, Gauging And Minkowski
  Compactification,''
  Phys.\ Lett.\ B {\bf 154}, 37 (1985).
  %%CITATION = PHLTA,B154,37;%%

\bibitem{AKK} T.~Gherghetta and A.~Kehagias,
  %``Anomaly free brane worlds in seven dimensions,''
  Phys.\ Rev.\ Lett.\  {\bf 90}, 101601 (2003)
  [arXiv:hep-th/0211019]; Phys.\ Rev.\ D {\bf 68}, 065019 (2003)
  [arXiv:hep-th/0212060].

  \bibitem{AKK2} S.~D.~Avramis and A.~Kehagias,
  %``Gauged D = 7 supergravity on the S(1)/Z(2) orbifold,''
  Phys.\ Rev.\ D {\bf 71}, 066005 (2005)
  [arXiv:hep-th/0407221].


%\cite{Bergshoeff:1985mr}
\bibitem{Bergshoeff:1985mr}
  E.~Bergshoeff, I.~G.~Koh and E.~Sezgin,
  ``Yang-Mills / Einstein Supergravity In Seven-Dimensions,''
  Phys.\ Rev.\ D {\bf 32}, 1353 (1985).
  %%CITATION = PHRVA,D32,1353;%%

%\cite{Kehagias:2004fb}
\bibitem{Kehagias:2004fb}
  A.~Kehagias,
  ``A conical tear drop as a vacuum-energy drain for the solution of the
  cosmological constant problem,''
  Phys.\ Lett.\ B {\bf 600}, 133 (2004)
  [arXiv:hep-th/0406025].
  %%CITATION = HEP-TH 0406025;%%

\bibitem{Nair:2004yu}
  V.~P.~Nair and S.~Randjbar-Daemi,
  ``Nonsingular 4d-flat branes in six-dimensional supergravities,''
  JHEP {\bf 0503}, 049 (2005)
  [arXiv:hep-th/0408063].
  %%CITATION = HEP-TH 0408063;%%

\end{thebibliography}
\end{document}